\documentclass[aps,pra,preprint,groupedaddress,longbibliography]{revtex4-1}
\usepackage{amsmath}
\usepackage{graphicx}
\usepackage{bm}
\allowdisplaybreaks[2]
\begin{document}

\title{Quantum teleportation in terms of Dirac modes between accelerated observers}

\author{Yue Dai}
\affiliation{Beijing Computational Science Research Center, Beijing 100193, China}

\author{Yu Shi}
\email[]{yushi@fudan.edu.cn}
\affiliation{Department of Physics  \&  State Key Laboratory of Surface Physics,   Fudan University,\\ Shanghai 200433, China}
%\date{\today}

\date{  }

\begin{abstract}
%abstract

We have studied entanglement between two Dirac modes respectively observed by two independently  accelerated observers. Due to Unruh effect, the entanglement degrades, but  residual nonzero entanglement remains even when the accelerations of both observers are infinite. We have also investigated quantum teleportation using entangled Dirac modes, and calculated the fidelities in various cases. In general, the fidelity depends on the single-qubit state $\gamma|0\rangle +\delta|1\rangle$ to be teleported, as well as which Bell state to use.  In the limit that both  accelerations are infinite,  the average fidelity in using  $|\phi^\pm\rangle\equiv\frac{1}{2}(|00\rangle\pm |11\rangle) $ is  $\frac{3}{4}$, while that in using  $|\psi^\pm\rangle\equiv \frac{1}{2} (|01\rangle\pm |10\rangle) $ is $\frac{1}{2} + |\gamma|^2 |\delta|^2$.

\end{abstract}
\maketitle

\section{Introduction}

In recent years, researches on quantum information have entered the realm of relativity \cite{RevModPhys.76.93,terno2006introduction,martin2011relativistic,mann2012relativistic,ralph2012relativistic}. It has been widely acknowledged that there exist gravitational or accelerating effects on quantum entanglement, which is one of the key concepts in quantum information theory. Because of the Unruh effect \cite{PhysRevD.7.2850,PhysRevD.14.870,RevModPhys.80.787}, entanglement between field modes degrades when at least one of the modes is observed by an accelerated observer \cite{PhysRevA.76.062112,  PhysRevA.83.012111, PhysRevA.77.024302,  PhysRevA.74.032326, TORRESARENAS201993,PhysRevA.83.022314, PhysRevA.85.024301, PhysRevA.85.032302, PhysRevA.92.022334, xiao.mixed, Shamirzaie2012, khan2014non, PhysRevA.92.022334,PhysRevLett.95.120404}.
Entanglement between two Dirac modes when one observer accelerates while the other moves uniformly was studied, indicating that the entanglement degrades with the acceleration. However, in contrast to the boson modes, whose entanglement approaches zero in the infinite acceleration limit \cite{PhysRevLett.95.120404, PhysRevA.76.062112, PhysRevA.83.012111, PhysRevA.77.024302, xiao.mixed, Shamirzaie2012, khan2014non, PhysRevA.92.022334}, the Dirac modes have finite residual entanglement in this limit \cite{PhysRevA.74.032326, PhysRevA.77.024302, PhysRevA.83.022314,PhysRevA.85.024301,TORRESARENAS201993, xiao.mixed, Shamirzaie2012, khan2014non,PhysRevA.85.032302, PhysRevA.92.022334}. The case that  two observers have the same acceleration was also studied, and by using the Bell states as the initial states, it was found that
entanglement degradation depends on  both the particle statistics and the initial state~\cite{PhysRevA.92.022334}.   With the increase of acceleration, the entanglement degradation of $|\psi^\pm\rangle\equiv (|01\rangle \pm |10\rangle)/\sqrt{2}$ is  faster than $|\phi^\pm\rangle \equiv (|00\rangle \pm |11\rangle)/\sqrt{2}$.  Entanglement among three Dirac modes with one or two observers accelerating has also been studied~\cite{PhysRevA.83.022314}. Entanglement between Unruh-Wald detectors of boson fields have also been studied and it was found that when one or more detectors accelerate, entanglement quickly degrades, and suddenly dies at  finite values  of the accelerations  \cite{PhysRevA.80.032315, Dai2015, PhysRevD.94.025012}.

Entanglement degradation leads to the reduction of the fidelity of the quantum teleportation~\cite{PhysRevLett.91.180404, alsing.teleportation,PhysRevA.80.032315,lidaishi}.
Therefore, for quantum teleportation between accelerated observers, it is better to use Dirac modes rather than boson modes.

In this paper, assuming both observers accelerate independently, we first  calculate the entanglement between two Dirac  modes in two generic kinds of entanglement state, using negativity as the entanglement measure.  Then we study  quantum teleportation using Dirac modes, and discuss the fidelities in various cases. We focus on how the fidelity depends on the accelerations of the observers. In Sec. \ref{secunruh}, we  briefly review the Unruh effect in order to summarize the formulae needed. In Sec. \ref{secentanglement}, we calculate the  entanglement degradation of two different  entangled states  of Dirac modes. Teleportation and its fidelities in  various cases are discussed in Sec. \ref{secfidelity}.   Summary and  discussions are made in Sec. \ref{seccon}.

\section{The Unruh effect}\label{secunruh}

\begin{figure}
\centering
\includegraphics[width=0.5\textwidth]{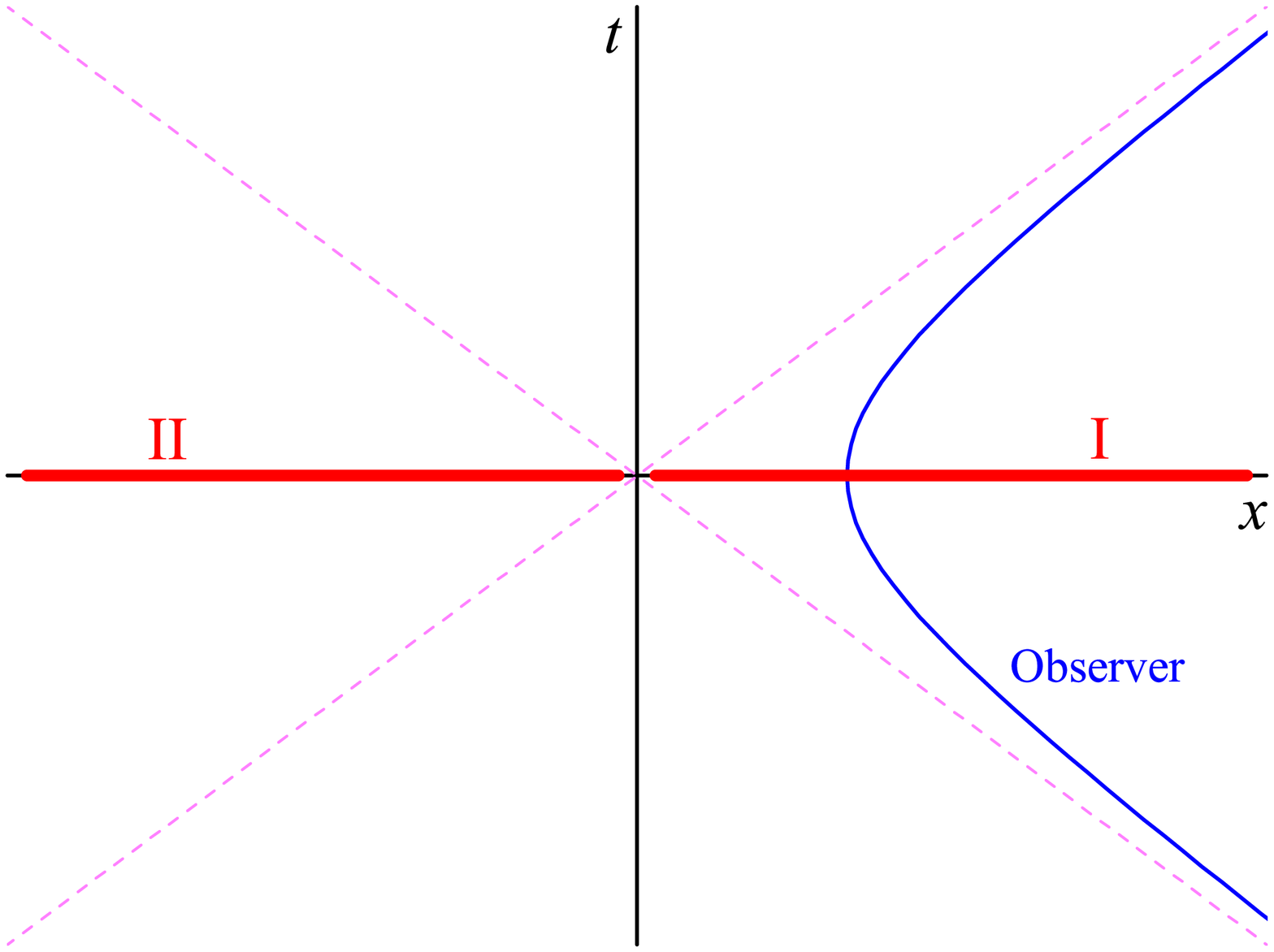}
\caption{\label{rindlerfig}} The Minkowski spacetime is divided into two parts,  the right Rindler wedge (region I), which can be fully accessed by the accelerating  observer, and the left Rindler wedge (region II), which is causally disconnected. The hyperbola   represents the world line of the observer accelerating uniformly in the $+x$ direction.
\end{figure}

In this section, we briefly review the Unruh effect in order to summarize  the formulae needed \cite{PhysRevA.74.032326,mcmahon2006dirac}.   For  simplicity, consider  a free Dirac field $\psi$,  satisfying, in  Minkowski spacetime,
\begin{equation}
(i\gamma^\mu \partial_\mu - m ) \psi = 0,
\end{equation}
where $m$ is the mass of the particle, $\gamma^\mu$'s  are  Dirac matrices. $\psi$ can be expanded in terms of the positive-frequency modes $\psi^+_k$ and the negative-frequency modes $\psi^-_k$, \cite{peskin1995quantum}
\begin{equation}\label{psieq1}
\psi = \int dk \left( a_k \psi^+_k + b^\dagger_k \psi^-_k\right),
\end{equation}
where we use wave vector $k$ to denote the modes, and  $a_k$ and $a^\dagger_k$ to denote the annihilation and creation operators of the positive-frequency  modes $
\psi ^{+}_k = \frac{1}{\sqrt{2\pi \omega_k}} \phi^+ (\mathbf{k}) e^{  i(kx-\omega_k t)}$, and $b_k$ and $b^\dagger_k$ to denote the annihilation and creation operators of   the negative-frequency modes $
\psi ^{-}_k =\frac{1}{\sqrt{2\pi \omega_k}}  \phi^-(\mathbf{k}) e^{\pm i(kx-\omega_k t)}$ with frequency $\pm\omega_k$.  $\phi^\pm(\mathbf{k}) $ is some constant spinor. The Minkowski vacuum is defined as
\begin{equation}
 |0\rangle = \prod_{kk'} |0_k\rangle^+ |0_{k'}\rangle^-,
\end{equation}
where $|0_k\rangle^+$ and $|0_{k'}\rangle^-$ are  fermion and antifermion vacua, satisfying  $a_k |0_k\rangle^+ = 0$ and $b_k |0_k\rangle^- = 0$ respectively.  One-fermion  state   can be defined as $|1_k\rangle^+ = a_k^\dag |0_k\rangle^+  $, and the one-antifermion   state  as  $|1_k\rangle^- = b_k^\dag |0_k\rangle^- $. The field operator $\psi$ is defined over the whole Minkowski spacetime. However, for an accelerating observer, only a part of the spacetime can be accessed \cite{mtw}. Consider an observer accelerating  in the $+x$ direction. The Minkowski spacetime  is divided into  the right Rindler wedge  I, and the left Rindler wedge  II (Fig. \ref{rindlerfig}). Region I can be fully accessed by the observer, while region II is causally disconnected.

One can also expand $\psi$ in terms of the Rindler modes in the two regions,
\begin{equation}\label{psieq2}
\psi = \int dk \left( c^{\rm{I}}_k \psi^{\rm{I}+}_k + d^{\rm{I}\dagger}_k \psi^{\rm{I}-}_k + c^{\rm{II}}_k \psi^{\rm{II}+}_k + d^{\rm{II}\dagger}_k \psi^{\rm{II}-}_k \right),
\end{equation}
where $\sigma\in \{ \rm{I},\rm{II}\}$, $c^{\sigma}_k$ is the annihilation operator  of the  fermion mode $\psi^{\sigma +}_k$, $d^{\sigma\dagger}_k$  is the creation  operator of the  antifermion mode $\psi^{\sigma -}_k$.

The vacua for fermions and antifermions in each Rindler wedge are defined through $c^\sigma_k |0_k\rangle^+_\sigma =0$ and $d^\sigma_k |0_k\rangle^-_\sigma =0$, respectively. The one-fermion and one-antifermion states are defined as $|1_k\rangle ^+_\sigma = c^{\sigma\dag}_k |0_k\rangle^+_\sigma$ and $|1_k\rangle ^-_\sigma = d^{\sigma\dag}_k |0_k\rangle^-_\sigma$, respectively.   From Eqs. (\ref{psieq1}) and (\ref{psieq2}), we find
\begin{eqnarray}
c^\sigma_k &=& \int dk' \left( \alpha^\sigma_{kk'}a_{k'} + \beta^\sigma_{kk'}b^\dagger_{k'} \right),\\
d^\sigma_k &=& \int dk' \left( \alpha^\sigma_{kk'}b_{k'} + \beta^\sigma_{kk'}a^\dagger_{k'} \right),
\end{eqnarray}
where $
\alpha^\sigma_{kk'} = \left( \psi^{\sigma+}_k, \psi^+_{k'} \right)$, $\beta^\sigma_{kk'} = \left( \psi^{\sigma+}_k, \psi^-_{k'}\right)$.  Under the single-mode approximation~\cite{PhysRevLett.91.180404,
alsing.teleportation},
one obtains  the  Bogoliubov transformation of the form  \cite{takagi1986vacuum, PhysRevD.43.3979, PhysRevA.74.032326,mcmahon2006dirac}
\begin{equation}
\left[ \begin{array}{l}
{a_k}\\
b_{ - k}^\dag
\end{array} \right] = \left[ {\begin{array}{*{20}{c}}
{\cos u}&{ - {e^{ - i\phi }}\sin u}\\
{{e^{ -i \phi }}\sin u}&{\cos u}
\end{array}} \right]\left[ \begin{array}{l}
c_k^{\rm{I}}\\
d_{ - k}^{{\rm{II}}\dag }
\end{array} \right].
\end{equation}
where $u$ is related to the acceleration $a$  through  \begin{equation}
\cos u = \left( e^{ - 2\pi \omega_k /a} + 1 \right)^{- 1/2},
 \label{ua}
 \end{equation}
hence $u$ is a measure  of the acceleration,  ranging  from $0$ to $\pi/4$,  $\phi$ is an  unimportant  phase  and can always be absorbed into the redefinition of the operators, and is disregarded henceforth.

Using the Bogoliubov transformation, we can express the Minkowski states in terms of the Rindler states. The Minkowski  vacuum can be  written as \cite{PhysRevA.74.032326}
$$\cos u \left| 0_k \right\rangle_{\rm{I}}^{+} \left| 0_{-k} \right\rangle_{\rm{II}}^{-} + \sin u\left| 1_k \right\rangle_{\rm{I}}^+ \left| 1_{-k} \right\rangle_{\rm{II}}^{-} ,$$
and the Minkowski one-fermion state can be written as
$$ \left| 1_k \right\rangle_{\rm{I}}^{+} \left| 0_{-k} \right\rangle_{\rm{II}}^{-}.  $$
The superscripts $+$ and $-$ represent the particle and antiparticle states, respectively, and are omitted henceforth, as $+$  is always associated with region I, and $-$ to region II.

The states of the modes probed by the accelerated observer can be obtained by tracing over the inaccessible region. Thus an accelerated observer in the Minkowski vacuum would probe a thermal state. This is the Unruh effect.

\section{Entanglement}\label{secentanglement}

We consider two entangled states of two modes  $A$ and $B$ in  Minkowski spacetime,
\begin{eqnarray}
  \left| \Psi  \right\rangle &=& \alpha \left| 0 \right\rangle^A \left| 1 \right\rangle^B + \beta \left| 1 \right\rangle^A \left| 0 \right\rangle^B ,\\
  \left| \Phi  \right\rangle &=& \alpha \left| 0 \right\rangle^A \left| 0 \right\rangle^B  + \beta \left| 1 \right\rangle^A \left| 1 \right\rangle^B.
\end{eqnarray}
The cases with $\alpha=\frac{1}{\sqrt{2}}$ and $\beta = \pm\frac{1}{\sqrt{2}}$ correspond the four Bell states. Now, we distribute mode $A$ to Alice, and $B$ to Bob. Alice and Bob are accelerated observers, with acceleration measures $u_a$ and $u_b$, respectively. Each observer can only access part of the Hilbert space, hence the entanglement is degraded. Note that even though the two observers cannot directly send and receive signals to and from each other because their  accelerations are different, they can indirectly do so through a third party.

\subsection{  $ \left| \Psi  \right\rangle $}

Consider the Minkowski state $ \left| \Psi  \right\rangle $. Wherever we omit labels $A$ and $B$,  the first mode is distributed to Alice, while the second to Bob. We can rewrite $\left| \Psi  \right\rangle$  in terms of the Rindler modes,
\begin{equation}\label{state1a}
\left| \Psi  \right\rangle  = \alpha \cos {u_a} \left| {01} \right\rangle _{\rm{I}}\left| {00} \right\rangle _{\rm{II}} + \alpha \sin {u_a}\left| {11} \right\rangle _{\rm{I}}\left| {10} \right\rangle _{\rm{II}}
 + \beta \cos {u_b}\left| {10} \right\rangle _{\rm{I}}\left| {00} \right\rangle _{\rm{II}} + \beta \sin {u_b}\left| {11} \right\rangle _{\rm{I}}\left| {01} \right\rangle _{\rm{II}},
\end{equation}
where $u_a$ and $u_b$ are measures of the acceleration of Alice and Bob, respectively, as given in Eq. (\ref{ua}). Alice and Bob are causally disconnected from region II, which should be traced over. The reduced density matrix is
\begin{equation}\label{rho1}
  \rho^{\rm{I}}_{\Psi} = F_1 \left| {01} \right\rangle \left\langle {01} \right| + I_1 \left| {01} \right\rangle \left\langle {10} \right| + {I^*_1 }\left| {10} \right\rangle \left\langle {01} \right| + G_1 \left| {10} \right\rangle \left\langle {10} \right| + H_1 \left| {11} \right\rangle \left\langle {11} \right|,
\end{equation}
where  subscript I on RHS has been omitted,
%\begin{subequations}
%\begin{eqnarray}
$F_1 = \left| \alpha \right|^2\cos ^2 u_a$, $
G_1 =\left| \beta  \right|^2\cos ^2 u_b$,
$H_1 =\left| \alpha \right|^2\sin ^2u_a + \left| \beta  \right|^2 \sin ^2 u_b$,
$I_1 =\alpha \beta ^* \cos u_a \cos u_b$.
%\end{eqnarray}
%\end{subequations}

\begin{figure}
\centering
\includegraphics[width=0.5\textwidth]{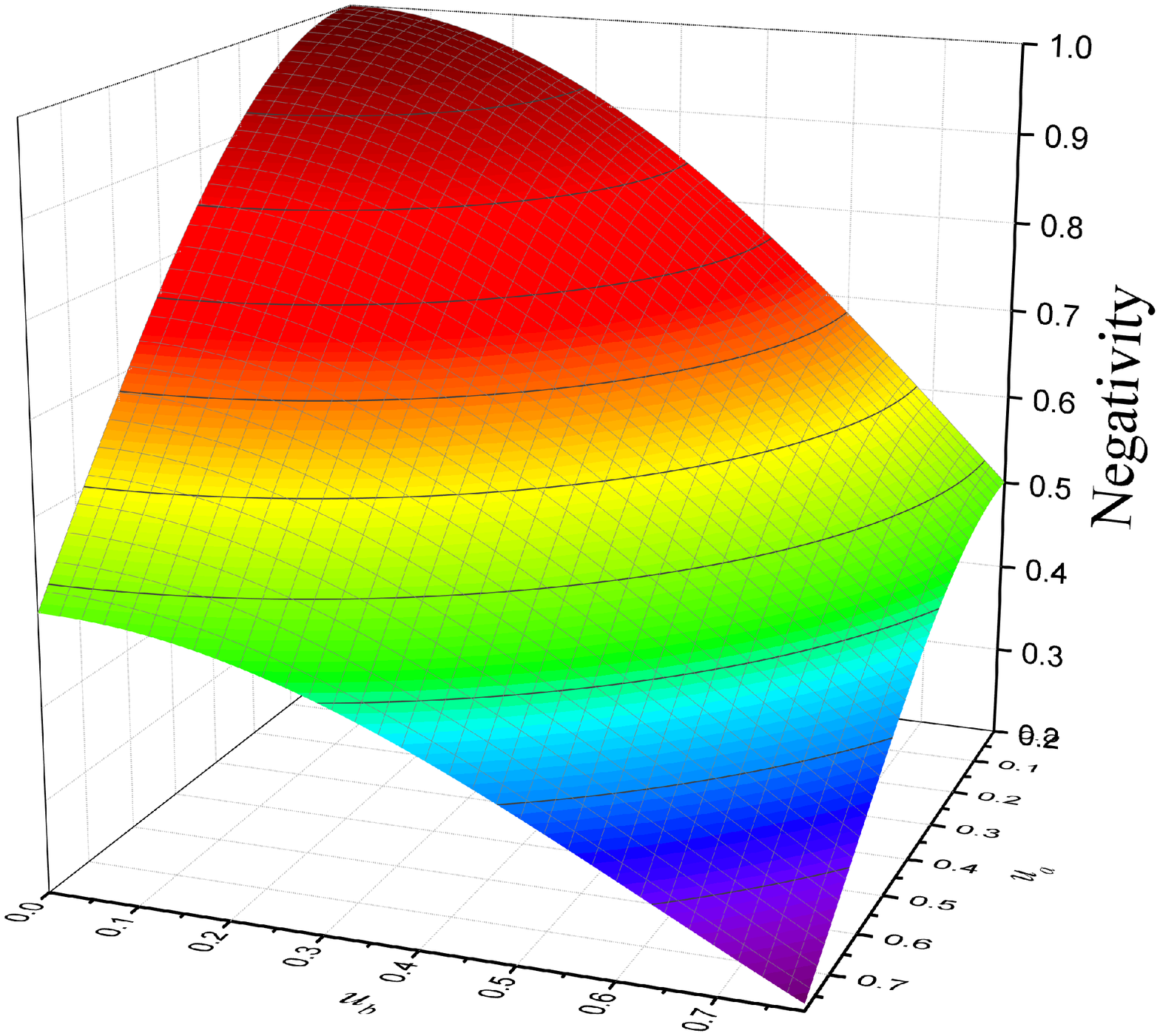}
\caption{\label{neg1}} The dependence of the negativity on $u_a$ and $u_b$ for $|\psi^\pm\rangle = \frac{1}{\sqrt{2}} \left(\left| {01} \right\rangle  \pm \left| {10} \right\rangle\right)$.
\end{figure}

To quantify the entanglement, we employ the   negativity
\begin{equation}
{\cal N} \equiv  \left\| \rho ^{T} \right\| - 1,
\end{equation}
where $T$ denotes a partial transpose with respect to $A$ or $B$, and $\|\rho\| \equiv \rm{Tr} \sqrt{\rho \rho^\dag}$ is the trace norm of $\rho$. The result is
\begin{equation}\label{negeq}
{\cal N} = \sum\limits_{i = 1}^4 {{\lambda _i}}  - 1.
\end{equation}
The squares of  $\lambda _i$'s are
$  {F_1^2}$, ${G_1^2}$, $\frac{{2{{\left| I_1 \right|}^2} + {H_1^2} \pm H_1\sqrt {{H_1^2} + 4{{\left| I_1 \right|}^2}} }}{2}$. In Fig.~\ref{neg1}, we show the dependence of the negativity on $u_a$ and $u_b$, for $|\alpha| = |\beta| = \frac{1}{\sqrt{2}}$.

The negativity has nonzero value in the limit of infinite acceleration, as a feature   of the entangled Dirac modes. When the acceleration of  Bob approaches infinity while Alice moves uniformly, the negativity approaches  $\left| \alpha  \right|^2 + \frac{1}{2}\left| \beta  \right|^2 + \frac{1}{2}\sqrt {8{\left| \alpha  \right|^2}\left| \beta  \right|^2 + \left| \beta  \right|^4}  - 1$. Especially, $\mathcal{N}\to \frac{1}{2}$  for $|\alpha|=|\beta|=\frac{1}{\sqrt{2}}$. When the accelerations of both observers approach infinity, the negativity approaches  $\frac{\sqrt {1 + 4\left| \alpha  \right|^2 \left| \beta  \right|^2}  - 1}{2}$, especially, $\mathcal{N}\to\frac{\sqrt{2}-1}{2}\approx 0.207$ for $|\alpha|=|\beta|=\frac{1}{\sqrt{2}}$.

\subsection{ $\left| \Phi  \right\rangle $}

\begin{figure}
\centering
\includegraphics[width=0.5\textwidth]{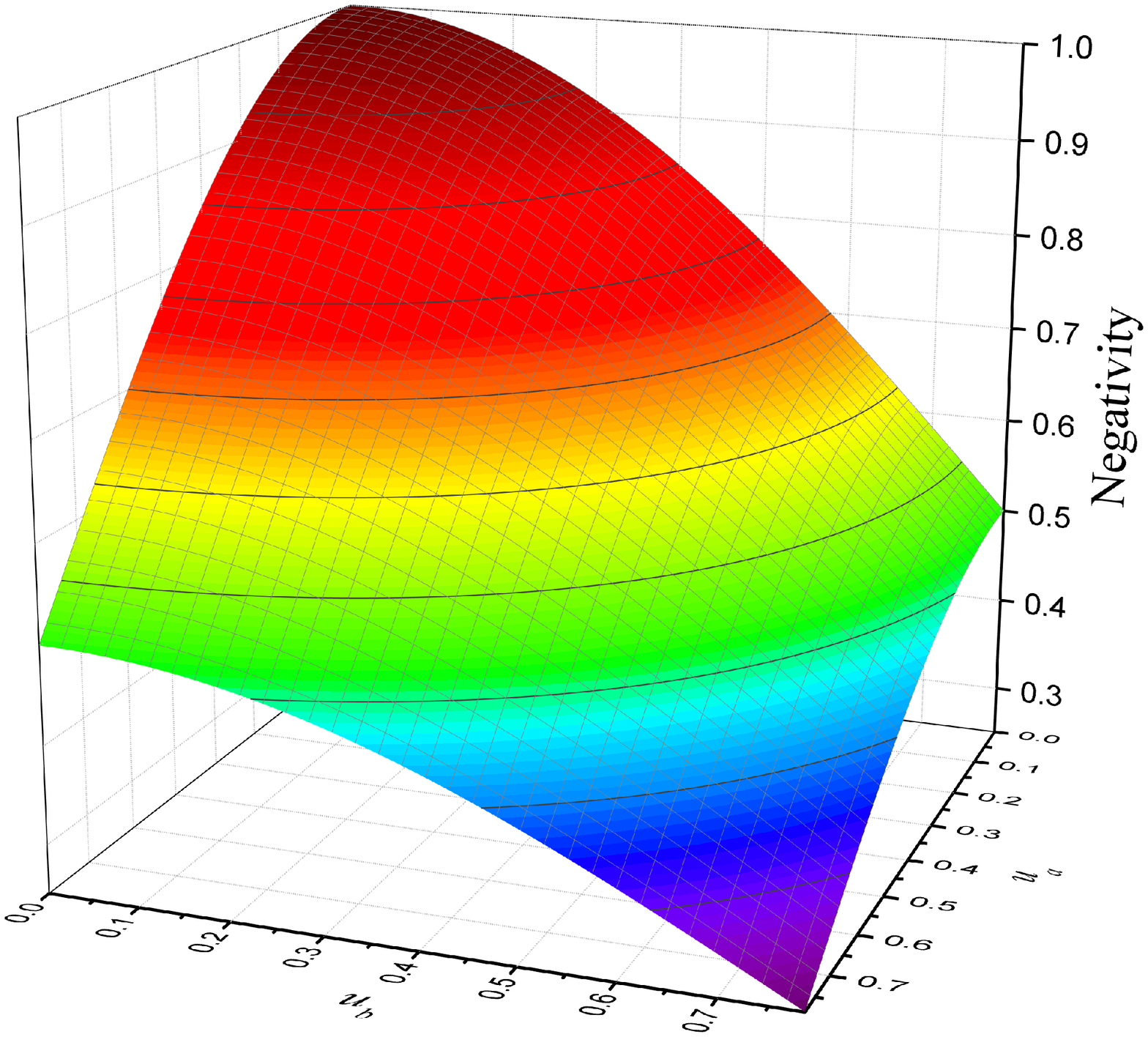}
\caption{\label{neg2}} The dependence of the negativity on $u_a$ and $u_b$ for $\left| \phi^\pm  \right\rangle = \frac{1}{\sqrt{2}} \left( \left| {00} \right\rangle  \pm \left| {11} \right\rangle\right)$.
\end{figure}

Now we  consider the Minkowski state $
  \left| \Phi  \right\rangle$,
which can be  rewritten as
\begin{eqnarray}\label{state2a}
\left| \Phi  \right\rangle  &=& \alpha \cos {u_a}\cos {u_b} \left| 00 \right\rangle _{\rm{I}} \left| 00 \right\rangle _{\rm{II}} + \alpha \cos {u_a}\sin {u_b} \left| 01 \right\rangle _{\rm{I}} \left| 01 \right\rangle _{\rm{II}}\nonumber\\
 &+& \alpha \sin {u_a}\cos {u_b}\left| 10 \right\rangle _{\rm{I}}\left| 10 \right\rangle _{\rm{II}} + \alpha \sin {u_a}\sin {u_b}\left| 11 \right\rangle _{\rm{I}}\left| 11 \right\rangle _{\rm{II}} + \beta \left| 11 \right\rangle _{\rm{I}}\left| 00 \right\rangle _{\rm{II}}
\end{eqnarray}
for the two accelerated observers. After tracing over region II, we obtain the reduced density matrix
\begin{equation}\label{rho2}
\rho^{\rm{I}}_{\Phi}  = F_2\left| 00 \right\rangle \left\langle 00 \right| + G_2\left| 01 \right\rangle \left\langle 01 \right| + H_2 \left| 10 \right\rangle \left\langle 10 \right| + I_2 \left| 11 \right\rangle \left\langle 11 \right| + J_2\left| 00 \right\rangle \left\langle 11 \right| + J_2^*\left| 11 \right\rangle \left\langle 00 \right|,
\end{equation}
where $
 F_2 = \left| \alpha  \right|^2 \cos ^2 {u_a} \cos ^2 {u_b}$,
 $G_2 = \left| \alpha  \right|^2 \cos ^2 {u_a} \sin ^2 {u_b}$,
 $H_2 = \left| \alpha  \right|^2 \sin ^2 {u_a} \cos ^2 {u_b}$,
 $I_2 = \left| \alpha  \right|^2 \sin ^2 {u_a} \sin ^2 {u_b} + \left| \beta  \right|^2$,
$ J_2  = \alpha \beta ^* \cos {u_a} \cos {u_b}$.
The negativity is also of the form of Eq. (\ref{negeq}), but
the squares of  $\lambda _i$'s are   ${F_2^2}$, ${I_2^2}$, $\frac{1}{2}\left[ {{G_2^2} + {H_2^2} + 2{|J_2|^2} \pm \sqrt {{{\left( {{G_2^2} - {H_2^2}} \right)}^2} + 4{{\left( {G_2 + H_2} \right)}^2}{|J_2|^2}} } \right]$.
For $|\alpha| = |\beta| = \frac{1}{\sqrt{2}}$, the result  is shown in Fig.~\ref{neg2}.

When the acceleration of Bob approaches infinity, while Alice moves uniformly, the negativity $\mathcal{N}$ approaches $\frac{1}{2}\left| \alpha  \right|^2 + \left| \beta  \right|^2 + \frac{1}{2}\sqrt {\left| \alpha  \right|^4 + 8\left| \alpha  \right|^2{\left| \beta  \right|^2}}  - 1$. Especially, $\mathcal{N}$ approaches $\frac{1}{2}$ for $|\alpha|=|\beta|=\frac{1}{\sqrt{2}}$. When the accelerations of both two observers approach  infinity,  the negativity $\mathcal{N}$ approaches  $\frac{1}{2}\left| \alpha  \right|^2 + \left| \beta  \right|^2 + \frac{1}{2}\sqrt {\frac{1}{2}\left| \alpha  \right|^4 + 2\left| \alpha  \right|^2 \left| \beta  \right|^2 + \left| \frac{1}{2}\left| \alpha  \right|^4 - 2\left| \alpha  \right|^2 \left| \beta  \right|^2 \right|}  - 1$, especially, $\mathcal{N}$ approaches $\frac{1}{4}$  for $|\alpha|=|\beta|=\frac{1}{\sqrt{2}}$.

For $|\alpha|=|\beta|=\frac{1}{\sqrt{2}}$ and  in the limit that Bob accelerates infinitely while Alice moves uniformly,  the negativities of $|\Psi \rangle$ and $|\Phi\rangle $ approach the same value.  But in general, for  same  accelerations, the negativities of $|\Psi \rangle$ and $|\Phi \rangle$ are different quantitatively, though the qualitative features are similar.

\section{Teleportation Fidelity}\label{secfidelity}

Quantum teleportation is usually realized by using one of the four Bell states \cite{PhysRevLett.70.1895}, $|\psi^\pm \rangle = (|10\rangle \pm |01\rangle)/\sqrt{2}$, $|\phi^\pm \rangle = (|00\rangle \pm |11\rangle)/\sqrt{2}$. Consider a Bell state of two modes $A$ and $B$, together with a qubit $Q$ in state $|Q\rangle$.   $A$ and $Q$ are controlled by Alice, $B$ is controlled by Bob. Alice performs a Bell measurement on $A$ and $Q$, resulting in  one of the four Bell states. Alice then communicates  to Bob the classical information about which state  $A$ and $Q$ are in. Then Bob applies a corresponding unitary transformation $U$ on the state of  $B$, which then becomes $|Q\rangle$, originally possessed by $Q$.

In the ideal protocol of quantum teleportation, the qubit state can be teleportated faithfully by using any of the Bell states. There's no real difference in using different Bell states. This is not the case when the observers  accelerate, since the degradation of the  entanglement now depends on the initial state.

We consider the situation that both Alice and Bob accelerate, characterized by $u_a$ and $u_b$ respectively.

\subsection{Quantum teleportation using $\left| \psi ^ \pm  \right\rangle $ }

First we consider  quantum teleportation using   $
  \left| \psi ^ \pm  \right\rangle$. $
  \left| \psi ^ \pm  \right\rangle$  is prepared in the inertial frames.  Suppose
$  \left| Q \right\rangle  = \gamma \left| 0 \right\rangle  + \delta \left| 1 \right\rangle$ with $|\gamma|^2 + |\delta|^2 = 1$. The initial state of the  system can be written as
\begin{equation}
  \left| \psi ^{QAB}_{\rm{in}} \right\rangle  = \frac{1}{\sqrt{2}}\left[ \gamma \left| 0 \right\rangle  + \delta \left| 1 \right\rangle  \right] \left[ \left| 10 \right\rangle \pm \left| 01 \right\rangle  \right].
\end{equation}
Here and in the following,  the upper and lower symbols in `$\pm$' or `$\mp$' correspond to the upper and lower symbols in the initial state $|\psi^\pm\rangle$ or $|\phi^\pm\rangle$, respectively.

As  Alice and Bob both accelerate,  the state $ \left| \psi ^{QAB}_{\rm{in}} \right\rangle $ can be rewritten  in terms of the Rindler modes,
\begin{eqnarray}\label{tele1}
\left| \psi ^{QAB}_{\rm{in}} \right\rangle  &=& \frac{1}{\sqrt{2}} \left[ \gamma \cos {u_b} \left| 010 \right\rangle _{\rm{I}} \left| 00 \right\rangle _{\rm{II}} + \gamma \sin {u_b}  \left| 011 \right\rangle_{\rm{I}} \left| 01 \right\rangle_{\rm{II}}\right.\nonumber\\
 &\pm& \gamma \cos {u_a}  \left| 001 \right\rangle_{\rm{I}} \left| 00 \right\rangle_{\rm{II}} \pm \gamma \sin {u_a}  \left| 011 \right\rangle_{\rm{I}}\left| 10 \right\rangle_{\rm{II}}\nonumber\\
 &+& \delta \cos {u_b} \left| 110 \right\rangle_{\rm{I}} \left| 00 \right\rangle _{\rm{II}} + \delta \sin {u_b}  \left| 111 \right\rangle_{\rm{I}} \left| 01 \right\rangle_{\rm{II}} \nonumber\\
  &\pm& \left. \delta \cos {u_a}  \left| 101 \right\rangle_{\rm{I}} \left| 00 \right\rangle_{\rm{II}} \pm \delta \sin {u_a} \left| 111 \right\rangle_{\rm{I}} \left| 10 \right\rangle_{\rm{II}} \right].
\end{eqnarray}
On RHS, the three numbers in each $|\ldots \rangle_{\rm{I}}$ consecutively  denote basis states of $Q$ and  the   Rindler modes of $A$ and $B$  in region I.  The two numbers in $|\ldots \rangle_{\rm{II}}$ consecutively denote basis states of the Rindler modes of $A$ and $B$ in region II. We rewrite Eq.~(\ref{tele1}) in terms of the Bell basis states $|\psi^\pm_{QA}\rangle$ and $|\phi^\pm_{QA}\rangle$ of the qubit and the Rindler mode of $A$ in region I,
\begin{eqnarray}
\left| \psi ^{QAB}_{\rm{in}} \right\rangle  &=& \frac{1}{2} \left[ \left| \psi^+_{QA}  \right\rangle \left( \gamma \cos {u_b} \left| 0 \right\rangle_{\rm{I}} \left| 00 \right\rangle_{\rm{II}} + \gamma \sin {u_b} \left| 1 \right\rangle_{\rm{I}} \left| 01 \right\rangle_{\rm{II}} \right. \right.\nonumber\\
 &\pm & \left. \gamma \sin {u_a} \left| 1 \right\rangle_{\rm{I}} \left| 10 \right\rangle_{\rm{II}} \pm \delta \cos {u_a} \left| 1 \right\rangle_{\rm{I}} \left| 00 \right\rangle_{\rm{II}} \right)\nonumber\\
 &+& \left| \psi^-_{QA}  \right\rangle \left( - \gamma \cos {u_b} \left| 0 \right\rangle_{\rm{I}} \left| 00 \right\rangle_{\rm{II}} - \gamma \sin {u_b} \left| 1 \right\rangle_{\rm{I}} \left| 01 \right\rangle_{\rm{II}} \right.\nonumber\\
 &\mp & \left. \gamma \sin {u_a} \left| 1 \right\rangle_{\rm{I}}\left| 10 \right\rangle_{\rm{II}} \pm \delta \cos {u_a} \left| 1 \right\rangle_{\rm{I}} \left| 00 \right\rangle_{\rm{II}} \right)\nonumber\\
 &+& \left| \phi^+_{QA} \right\rangle \left( \pm \gamma \cos {u_a} \left| 1 \right\rangle_{\rm{I}} \left| 00 \right\rangle_{\rm{II}} + \delta \cos {u_b} \left| 0 \right\rangle_{\rm{I}} \left| 00 \right\rangle_{\rm{II}} \right.\nonumber\\
 &+& \left. \delta \sin {u_b} \left| 1 \right\rangle_{\rm{I}} \left| 01 \right\rangle_{\rm{II}} \pm \delta \sin {u_a} \left| 1 \right\rangle_{\rm{I}} \left| 10 \right\rangle_{\rm{II}} \right)\nonumber\\
 &+& \left| \phi^-_{QA} \right\rangle \left( \mp\gamma \cos {u_a} \left| 1 \right\rangle_{\rm{I}} \left| 00 \right\rangle_{\rm{II}} + \delta \cos {u_b} \left| 0 \right\rangle_{\rm{I}} \left| 00 \right\rangle_{\rm{II}} \right.\nonumber\\
 &+& \left. \left. \delta \sin {u_b} \left| 1 \right\rangle_{\rm{I}} \left| 01 \right\rangle_{\rm{II}} \pm \delta \sin {u_a} \left| 1 \right\rangle_{\rm{I}} \left| 10 \right\rangle_{\rm{II}} \right) \right],
\end{eqnarray}
where  the number in each $|\ldots\rangle_{\rm{I}}$ denotes a basis state of the Rindler mode of $B$.

Now Alice makes a joint Bell measurement on $A$ and $Q$.    Because of Unruh effect, the state of Bob is now a mixed state. It is on this mixed state that  Bob makes a unitary transformation $U$ on $B$.

If the result is $|\psi^+_{QA}\rangle$, with probability $p_1\equiv(|\gamma|^2+ |\gamma|^2 \sin^2 u_a +|\delta|^2 \cos ^2 u_a)  /4$, the state of the system becomes
\begin{eqnarray}
\left| \psi_1^{QAB}  \right\rangle  &=& \frac{\rm{1}}{\eta_1} \left| \psi ^+_{QA}  \right\rangle \left( \gamma \cos {u_b} \left| 0 \right\rangle_{\rm{I}} \left| 00 \right\rangle_{\rm{II}} + \gamma \sin {u_b} \left| 1 \right\rangle_{\rm{I}} \left| 01 \right\rangle_{\rm{II}} \right. \nonumber\\
&& \left. \pm \gamma \sin {u_a} \left| 1 \right\rangle_{\rm{I}} \left| 10 \right\rangle_{\rm{II}} \pm \delta \cos {u_a} \left| 1 \right\rangle_{\rm{I}} \left| 00 \right\rangle_{\rm{II}} \right),
\end{eqnarray}
where $\frac{1}{\eta_1}$ is the normalization factor, $  \eta_1^2 = | \delta |^2 \cos^2 {u_a} + | \gamma |^2 \left( 1 + \sin^2 {u_a} \right)$.
The reduced density matrix  observed by  Bob,  obtained by tracing over the qubit, the Rindler mode  of $A$ in I,  as well as region II, is
\begin{eqnarray}
\rho_{B1} &=& \frac{1}{\eta_1^2} \left \{ | \gamma |^2\cos^2 {u_b}  \left| 0 \right\rangle \left\langle 0 \right|\pm \gamma \delta^* \cos {u_a} \cos {u_b} \left| 0 \right\rangle \left\langle 1 \right| \right.\nonumber\\
&& \left.\pm  \gamma^* \delta \cos {u_a} \cos {u_b}  \left| 1 \right\rangle \left\langle 0 \right| + \left[ | \gamma |^2 \left( \sin^2{u_a} + \sin^2 {u_b} \right)  + | \delta |^2 \cos^2 {u_a} \right]\left| 1 \right\rangle \left\langle 1 \right| \right\},
\end{eqnarray}
where the subscript  I is omitted. Bob makes a corresponding unitary transformation
\begin{equation}
\rho_{B1}   \longrightarrow U^\dag \rho_{B1} U,
\end{equation}
where $U =  |0\rangle\langle 0| \pm |1\rangle\langle 1|$.  Thus the fidelity is
\begin{eqnarray}
 {\cal F}_1 &=& \langle Q | U^\dag \rho_{B1} U | Q \rangle \nonumber\\
 &=& \frac{1}{\eta_1^2} \left[ \left( | \gamma |^2 \cos {u_b}  + | \delta |^2 \cos {u_a} \right)^2 + | \gamma  |^2 | \delta |^2 \left( \sin^2 {u_a} + \sin^2 {u_b} \right)  \right],
\end{eqnarray}
The fidelity depends on the state $|Q\rangle$ teleported  unless  $u_a=u_b=0$, in which case ${\cal F}_1=1$.

If the result of the Bell measurement is $|\psi^-_{QA}\rangle$, with probability $p_1\equiv(|\gamma|^2+ |\gamma|^2 \sin^2 u_a +|\delta|^2 \cos ^2 u_a)  /4$, the state of the system becomes
\begin{eqnarray}
\left| {\psi_1^{QAB}}'  \right\rangle  &=& \frac{\rm{1}}{\eta_1} \left| \psi ^-_{QA}  \right\rangle \left( -\gamma \cos {u_b} \left| 0 \right\rangle_{\rm{I}} \left| 00 \right\rangle_{\rm{II}} - \gamma \sin {u_b} \left| 1 \right\rangle_{\rm{I}} \left| 01 \right\rangle_{\rm{II}} \right. \nonumber\\
&& \left. \mp \gamma \sin {u_a} \left| 1 \right\rangle_{\rm{I}} \left| 10 \right\rangle_{\rm{II}} \pm \delta \cos {u_a} \left| 1 \right\rangle_{\rm{I}} \left| 00 \right\rangle_{\rm{II}} \right).
\end{eqnarray}
The reduced density matrix  observed by  Bob  is then
\begin{eqnarray}
\rho_{B1}' &=& \frac{1}{\eta_1^2} \left \{ | \gamma |^2\cos^2 {u_b}  \left| 0 \right\rangle \left\langle 0 \right|\mp  \gamma \delta^* \cos {u_a} \cos {u_b} \left| 0 \right\rangle \left\langle 1 \right| \right.\nonumber\\
&& \left.\mp \gamma^* \delta \cos {u_a} \cos {u_b}  \left| 1 \right\rangle \left\langle 0 \right| + \left[ | \gamma |^2 \left( \sin^2{u_a} + \sin^2 {u_b} \right)  + | \delta |^2 \cos^2 {u_a} \right]\left| 1 \right\rangle \left\langle 1 \right| \right\},
\end{eqnarray}
where the subscript  I is omitted. Then on $\rho_{B1}'$, Bob makes  unitary transformation  $U = -|0\rangle\langle 0| \pm   |1\rangle\langle 1|$.
The fidelity is
\begin{eqnarray}
 {\cal F}_1' &=& \langle Q | U^\dag \rho_{B1} U | Q \rangle ={\cal F}_1.
\end{eqnarray}

If the result of the Bell measurement by Alice is $|\phi^+_{QA}\rangle$, each with probability $p_2\equiv(|\delta|^2+ |\delta|^2 \sin^2 u_a +|\gamma|^2 \cos ^2 u_a)  /4$,   the state of whole system   becomes
\begin{eqnarray}
\left| \psi^{QAB}_2  \right\rangle &=& \frac{1}{\eta_2} \left| \phi^\pm_{QA} \right\rangle \left( \pm \gamma \cos {u_a} \left| 1 \right\rangle_{\rm{I}} \left| 00 \right\rangle_{\rm{II}} + \delta \cos {u_b} \left| 0 \right\rangle_{\rm{I}} \left| 00 \right\rangle_{\rm{II}} \right.\nonumber\\
&& \left. + \delta \sin {u_b} \left| 1 \right\rangle_{\rm{I}} \left| 01 \right\rangle_{\rm{II}} \pm\delta \sin {u_a} \left| 1 \right\rangle_{\rm{I}} \left| 10 \right\rangle_{\rm{II}} \right),
\end{eqnarray}
where $\frac{1}{\eta_2}$ is the normalization factor, $
  \eta_2^2 = | \gamma |^2 \cos^2 {u_a}  + | \delta |^2 \left( 1 + \sin^2 {u_a} \right)$.
The reduced density matrix observed by Bob is then
\begin{eqnarray}
\rho_{B_2} &=& \frac{1}{\eta_2^2} \left\{ | \delta |^2 \cos^2 {u_b} | 0 \rangle \langle 0 | \pm \gamma^* \delta \cos {u_a} \cos {u_b} | 0 \rangle \langle 1 | \right.\nonumber\\
&&\left. { \pm \gamma \delta^* \cos {u_a} \cos {u_b} | 1 \rangle \langle 0 | + \left[ | \gamma |^2 \cos^2 {u_a} + | \delta |^2 \left( \sin^2 {u_a} + \sin^2 {u_b} \right) \right] | 1 \rangle \langle 1 |} \right\},
\end{eqnarray}
on which Bob makes  unitary transformation $U =  \pm |0\rangle\langle 1| +|1\rangle\langle 0|$.
Thus the fidelity is
\begin{eqnarray}\label{fidres2}
 {\cal F}_2 &=& \langle Q | U^\dag \rho_{B2} U | Q \rangle \nonumber\\
 &=& \frac{1}{\eta_2^2} \left[ \left( | \gamma |^2 \cos {u_a}  + | \delta |^2 \cos {u_b} \right)^2 + | \gamma |^2 | \delta |^2 \left( \sin^2 {u_a} + \sin^2 {u_b} \right) \right],
\end{eqnarray}

If the result of the Bell measurement by Alice is  $|\phi^-_{QA}\rangle$, the state of the system becomes
\begin{eqnarray}
\left| {\psi^{QAB}_2}'  \right\rangle &=& \frac{1}{\eta_2} \left| \phi^-_{QA} \right\rangle \left( \mp \gamma \cos {u_a} \left| 1 \right\rangle_{\rm{I}} \left| 00 \right\rangle_{\rm{II}} + \delta \cos {u_b} \left| 0 \right\rangle_{\rm{I}} \left| 00 \right\rangle_{\rm{II}} \right.\nonumber\\
&& \left. + \delta \sin {u_b} \left| 1 \right\rangle_{\rm{I}} \left| 01 \right\rangle_{\rm{II}} \pm  \delta \sin {u_a} \left| 1 \right\rangle_{\rm{I}} \left| 10 \right\rangle_{\rm{II}} \right).
\end{eqnarray}
The reduced density matrix observed by Bob is
\begin{eqnarray}
\rho'_{B_2} &=& \frac{1}{\eta_2^2} \left\{ | \delta |^2 \cos^2 {u_b} | 0 \rangle \langle 0 | \mp \gamma^* \delta \cos {u_a} \cos {u_b} | 0 \rangle \langle 1 | \right.\nonumber\\
&&\left. { \mp \gamma \delta^* \cos {u_a} \cos {u_b} | 1 \rangle \langle 0 | + \left[ | \gamma |^2 \cos^2 {u_a} + | \delta |^2 \left( \sin^2 {u_a} + \sin^2 {u_b} \right) \right] | 1 \rangle \langle 1 |} \right\},
\end{eqnarray}
on which Bob makes unitary transformation
$U = \mp |0\rangle\langle 1| +|1\rangle\langle 0|$. Thus the fidelity is
\begin{eqnarray}\label{fidres1}
 {\cal F}_2' &=& \langle Q | U^\dag \rho_{B1}' U | Q \rangle= {\cal F}_2.
\end{eqnarray}

Moreover, the average fidelity over all possibilities  is
\begin{eqnarray}
{\cal F}_{\psi}&=& 2p_1{\cal F}_1+2p_2{\cal F}_2\nonumber\\
%&=&\frac{1}{2}\left[ (|\gamma|^4 +|\delta|^4)(\cos^2 u_a +\cos^2 u_b) + 2|\gamma|^2 |\delta|^2 (\sin^2 u_a + \sin^2 u_b +\cos u_a \cos u_b)\right]\\
&=&\frac{1}{2}\left[ \cos^2 u_a +\cos^2 u_b \right. \nonumber \\
&& \left. + 2|\gamma|^2 |\delta|^2 (\sin^2 u_a -\cos^2 u_a + \sin^2 u_b -\cos^2 u_b  +2\cos u_a \cos u_b)\right],
\end{eqnarray}
which is symmetric between $\gamma$ and $\delta$.

\begin{figure}
\centering
\includegraphics[width=0.5\textwidth]{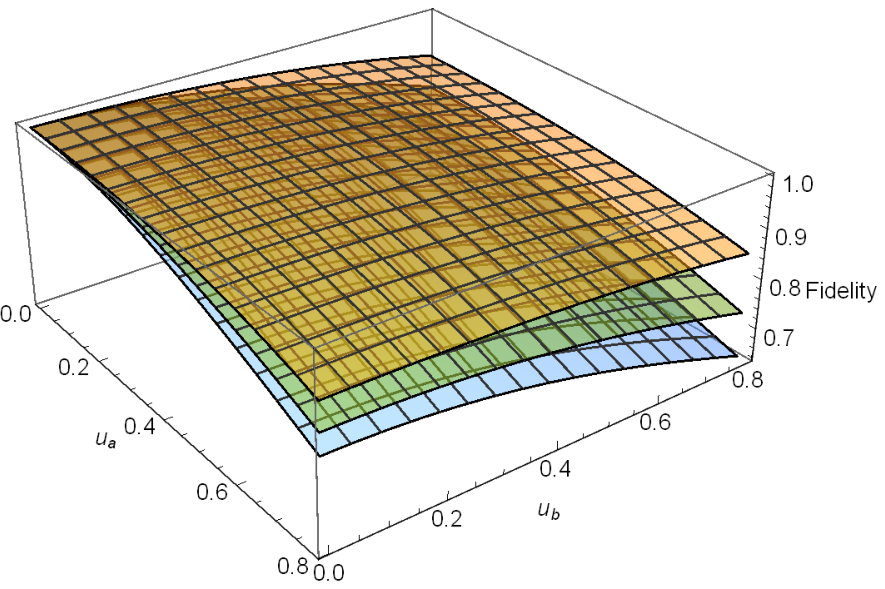}
\caption{\label{fid1}} In teleportation using $|\psi^\pm\rangle$, the fidelities ${\cal F}_1$ (the yellow layer),  ${\cal F}_2$ (the blue layer) and ${\cal F}_\psi$ (the green layer) as functions of $u_a$ and $u_b$. The coefficients characterizing  the teleported state are $|\gamma| = 0.6$ and $|\delta| = 0.8$.
\end{figure}

In the usual procedure of teleportation,  the fidelity do not depend on the result of Bell measurement. However, it is not the case when Alice and Bob accelerate. ${\cal F}_1 $ and ${\cal F}_2 $   are different in general.
When  $|\gamma| < |\delta|$, ${\cal F}_1> {\cal F}_2$, as indicated in  Fig.~\ref{fid1}, where  we show ${\cal F}_1$, ${\cal F}_2$ and ${\cal F}_\psi$ for $|\gamma| = 0.6$ and $|\delta| = 0.8$. When   $|\gamma| > |\delta|$, ${\cal F}_2 > {\cal F}_1$. In fact, one can obtain ${\cal F}_2$ from ${\cal F}_1$ by exchanging $\gamma$ and $\delta$.

We now consider the special case that $|\gamma|^2=|\delta|^2 = \frac{1}{2}$. In this special case,   ${\cal F}_1 = {\cal F}_2   = {\cal F}_\psi = \frac{1}{2}\left[ 1 + \cos {u_a}\cos {u_b} \right] $. In the limit that one observer's  acceleration is  infinity while the other moves uniformly, ${\cal F}_1  = {\cal F}_2   = {\cal F}_\psi= \frac{2+\sqrt{2}}{4}\approx 0.854$. When   the accelerations of   both observers  are infinity, ${\cal F}_1 = {\cal F}_2   = {\cal F}_\psi= \frac{3}{4}$.

\subsection{Quantum teleportation using $\left| \phi ^ \pm  \right\rangle  $}
Now we consider   using  $
  \left| \phi ^ \pm  \right\rangle $ to teleport
$|Q\rangle$. The initial state of the system is
\begin{equation}
  \left| \phi^{QAB}_{\rm{in}} \right\rangle  = \frac{1}{\sqrt{2}} \left[ \gamma | 0 \rangle + \delta | 1 \rangle \right] \left[ | 11 \rangle \pm | 00 \rangle \right],
\end{equation}
which can be expressed in terms of the Rindler modes as    \begin{eqnarray}\label{tele2}
\left| \phi^{QAB}_{\rm{in}} \right\rangle  &=& \frac{1}{\sqrt{2}} \left[ \gamma\cos {u_a} \cos {u_b} | 000 \rangle_{\rm{I}} | 00 \rangle_{\rm{II}} + \gamma \cos {u_a} \sin {u_b} | 001 \rangle_{\rm{I}} | 01 \rangle_{\rm{II}} \right. \nonumber\\
 &&+ \gamma \sin {u_a} \cos {u_b} | 010 \rangle_{\rm{I}} | 10 \rangle_{\rm{II}} + \gamma \sin {u_a} \sin {u_b} | 011 \rangle_{\rm{I}} | 11 \rangle_{\rm{II}} \nonumber\\
 &&+ \delta \cos {u_a} \cos {u_b} | 100 \rangle_{\rm{I}} | 00 \rangle_{\rm{II}} + \delta \cos {u_a} \sin {u_b} | 101 \rangle_{\rm{I}} | 01 \rangle_{\rm{II}} \nonumber\\
 &&+ \delta \sin {u_a} \cos {u_b} | 110 \rangle_{\rm{I}} | 10 \rangle_{\rm{II}} + \delta \sin {u_a} \sin {u_b} | 111 \rangle_{\rm{I}} | 11 \rangle_{\rm{II}} \nonumber\\
 &&\left.  \pm \gamma | 011 \rangle_{\rm{I}} | 00 \rangle_{\rm{II}} \pm \delta | 111 \rangle_{\rm{I}} | 00 \rangle_{\rm{II}} \right],
\end{eqnarray}
which can be rewritten in   in terms of Bell states of $Q$ and $A$,  as
\begin{eqnarray}
\left| \phi^{QAB}_{\rm{in}} \right\rangle &=& \frac{1}{2} \left[ | \psi^+_{QA} \rangle \left( \gamma \sin {u_a} \cos {u_b} | 0 \rangle_{\rm{I}} | 10 \rangle_{\rm{II}} + \gamma \sin {u_a} \sin {u_b} | 1 \rangle_{\rm{I}} | 11 \rangle_{\rm{II}} \right. \right.\nonumber\\
 &&+  \left. \delta \cos {u_a} \cos {u_b} | 0 \rangle_{\rm{I}} | 00 \rangle_{\rm{II}} + \delta \cos {u_a}\sin {u_b} | 1 \rangle_{\rm{I}} | 01 \rangle_{\rm{II}} \pm \gamma | 1 \rangle_{\rm{I}} | 00 \rangle_{\rm{II}} \right)\nonumber\\
 && + | \psi^-_{QA} \rangle \left( - \gamma \sin {u_a} \cos {u_b} | 0 \rangle_{\rm{I}} | 10 \rangle_{\rm{II}} - \gamma \sin {u_a} \sin {u_b} | 1 \rangle_{\rm{I}} | 11 \rangle_{\rm{II}} \right.\nonumber\\
 && + \left. \delta \cos {u_a} \cos {u_b} | 0 \rangle_{\rm{I}} | 00 \rangle_{\rm{II}} + \delta \cos {u_a} \sin {u_b} | 1 \rangle_{\rm{I}} | 01 \rangle_{\rm{II}} \mp \gamma | 1 \rangle_{\rm{I}}  | 00 \rangle_{\rm{II}} \right)\nonumber\\
 && + | \phi^+_{QA} \rangle \left( \gamma \cos {u_a}\cos {u_b} | 0 \rangle_{\rm{I}} | 00 \rangle_{\rm{II}} + \gamma \cos {u_a} \sin {u_b} | 1  \rangle_{\rm{I}} | 01 \rangle_{\rm{II}} \right.\nonumber\\
 && + \left.  \delta \sin {u_a} \cos {u_b} | 0 \rangle_{\rm{I}} | 10 \rangle_{\rm{II}} + \delta \sin {u_a} \sin {u_b} | 1 \rangle_{\rm{I}} | 11 \rangle_{\rm{II}} \pm \delta | 1 \rangle_{\rm{I}} | 00 \rangle_{\rm{II}} \right)\nonumber\\
 &&+ | \phi^-_{QA}  \rangle \left(  - \gamma \cos {u_a}\cos {u_b} | 0 \rangle_{\rm{I}} | 00 \rangle_{\rm{II}} - \gamma \cos {u_a} \sin {u_b} | 1 \rangle_{\rm{I}}  | 01 \rangle_{\rm{II}} \right.\nonumber\\
 && + \left. \left.  \delta \sin {u_a}\cos {u_b} | 0 \rangle_{\rm{I}} | 10 \rangle_{\rm{II}} + \delta  \sin {u_a} \sin {u_b} | 1  \rangle_{\rm{I}} | 11 \rangle_{\rm{II}} \pm \delta | 1 \rangle_{\rm{I}} | 00 \rangle_{\rm{II}} \right) \right].
\end{eqnarray}

Now Alice makes the Bell measurement.  If the result is $|\psi^+ \rangle$,   with probability $p_1\equiv(|\gamma|^2+ |\gamma|^2 \sin^2 u_a +|\delta|^2 \cos ^2 u_a)  /4$, the state  of the whole system becomes
\begin{eqnarray}
| \phi^{QAB}_3 \rangle &=& \frac{1}{\eta_1} | \psi^\pm_{QA} \rangle \left(  \gamma \sin {u_a} \cos {u_b} | 0 \rangle_{\rm{I}} | 10 \rangle_{\rm{II}} + \gamma \sin {u_a} \sin {u_b} | 1 \rangle_{\rm{I}} | 11 \rangle_{\rm{II}} \right.\nonumber\\
&& + \left.  \delta \cos {u_a} \cos {u_b} | 0 \rangle_{\rm{I}} | 00 \rangle_{\rm{II}} + \delta \cos {u_a} \sin {u_b} | 1 \rangle_{\rm{I}} | 01 \rangle_{\rm{II}} \pm \gamma  | 1 \rangle_{\rm{I}} | 00 \rangle_{\rm{II}} \right).
\end{eqnarray}
The reduced density matrix observed by Bob is
\begin{eqnarray}
\rho_{B3} &=& \frac{1}{\eta_1^2} \left[ \left( | \gamma |^2 \sin^2 {u_a} \cos^2 {u_a} + | \delta |^2 \cos^2 {u_a} \cos^2 {u_b}  \right) | 0 \rangle \langle 0 | \right.\nonumber\\
 &&\pm \gamma^* \delta \cos {u_a} \cos {u_b} | 0 \rangle \langle 1 | \pm \gamma \delta^* \cos {u_a} \cos {u_b} | 1 \rangle \langle 0 |\nonumber\\
 &&+ \left.  \left( | \gamma |^2\sin^2 {u_a} \sin^2 {u_b}  + | \delta |^2 \cos^2 {u_a} \sin^2 {u_b} + | \gamma  |^2 \right) | 1 \rangle \langle 1 | \right],
\end{eqnarray}
on which Bob makes unitary transformation  $U = \pm |0\rangle\langle 1| + |1\rangle\langle 0|$.
Thus we obtain the fidelity
\begin{eqnarray}
  {\cal F}_3 &=& \langle Q | U^\dag \rho_{B3} U | Q \rangle \nonumber\\
 &=& \frac{1}{\eta_1^2} \left[ | \gamma |^4 \left( 1 +  \sin^2 {u_a}\sin^2{u_b} \right) + | \delta |^4 \cos^2 {u_a}\cos ^2 {u_b} \right.\nonumber\\
 &&+ \left. | \gamma  |^2 | \delta |^2 \left( \sin^2 {u_a} \cos^2 {u_b} + \cos^2 {u_a} \sin^2 {u_b} + 2 \cos {u_a} \cos {u_b} \right) \right].
\end{eqnarray}

If the result of the Bell measurement is $|\psi^- \rangle$,   with probability $p_1\equiv(|\gamma|^2+ |\gamma|^2 \sin^2 u_a +|\delta|^2 \cos ^2 u_a)  /4$, the state  of the whole system becomes
\begin{eqnarray}
| {\phi^{QAB}_3}' \rangle &=& \frac{1}{\eta_1} | \psi^-_{QA} \rangle \left( - \gamma \sin {u_a} \cos {u_b} | 0 \rangle_{\rm{I}} | 10 \rangle_{\rm{II}} - \gamma \sin {u_a} \sin {u_b} | 1 \rangle_{\rm{I}} | 11 \rangle_{\rm{II}} \right.\nonumber\\
&& + \left.  \delta \cos {u_a} \cos {u_b} | 0 \rangle_{\rm{I}} | 00 \rangle_{\rm{II}} + \delta \cos {u_a} \sin {u_b} | 1 \rangle_{\rm{I}} | 01 \rangle_{\rm{II}} \mp  \gamma  | 1 \rangle_{\rm{I}} | 00 \rangle_{\rm{II}} \right),
\end{eqnarray}
The reduced density matrix observed by Bob is
\begin{eqnarray}
\rho_{B3}' &=& \frac{1}{\eta_1^2} \left[ \left( | \gamma |^2 \sin^2 {u_a} \cos^2 {u_a} + | \delta |^2 \cos^2 {u_a} \cos^2 {u_b}  \right) | 0 \rangle \langle 0 | \right.\nonumber\\
 &&\mp \gamma^* \delta \cos {u_a} \cos {u_b} | 0 \rangle \langle 1 | \mp \gamma \delta^* \cos {u_a} \cos {u_b} | 1 \rangle \langle 0 |\nonumber\\
 &&+ \left.  \left( | \gamma |^2\sin^2 {u_a} \sin^2 {u_b}  + | \delta |^2 \cos^2 {u_a} \sin^2 {u_b} + | \gamma  |^2 \right) | 1 \rangle \langle 1 | \right],
\end{eqnarray}
on which Bob makes unitary transformation $U = \mp |0\rangle\langle 1| + |1\rangle\langle 0|$.
Thus we obtain the fidelity
\begin{eqnarray}
  {\cal F}_3' &=& \langle Q | U^\dag \rho_{B3} U | Q \rangle=  {\cal F}_3.
\end{eqnarray}

If the Bell measurement results in  $|\phi^+\rangle$, with probability $p_2\equiv (|\delta|^2+ |\delta|^2 \sin^2 u_a +|\gamma|^2 \cos ^2 u_a)  /4$, the  state  of whole system is
\begin{eqnarray}
| \phi^{QAB}_4 \rangle &=& \frac{1}{\eta_2} | \phi^\pm_{QA} \rangle \left( \gamma \cos {u_a} \cos {u_b} | 0 \rangle_{\rm{I}} | 00 \rangle_{\rm{II}}+ \gamma \cos {u_a} \sin {u_b} | 1 \rangle_{\rm{I}} | 01 \rangle_{\rm{II}} \right.\nonumber\\
&&+ \left. \delta \sin {u_a} \cos {u_b} | 0 \rangle_{\rm{I}} | 10 \rangle_{\rm{II}} + \delta \sin {u_a} \sin {u_b} | 1 \rangle_{\rm{I}} | 11 \rangle_{\rm{II}} \pm  \delta | 1 \rangle_{\rm{I}} | 00 \rangle_{\rm{II}} \right).
\end{eqnarray}
The reduced density matrix observed by Bob is
\begin{eqnarray}
 \rho_{B4} &=& \frac{1}{\eta_2^2} \left[ \left( | \gamma |^2 \cos^2 {u_a} \cos^2 {u_b} + | \delta |^2 \sin^2 {u_a} \cos^2 {u_b} \right) | 0 \rangle \langle 0 | \right) \nonumber\\
 && \pm \gamma \delta^* \cos {u_a} \cos {u_b} | 0 \rangle \langle 1 | \pm \gamma ^* \delta \cos {u_a} \cos {u_b} | 1 \rangle \langle 0 |\nonumber\\
 && + \left. \left( | \gamma |^2 \cos^2 {u_a} \sin^2 {u_b} + | \delta |^2 \sin^2 {u_a} \sin^2 {u_b}  + | \delta |^2 \right) | 1 \rangle \langle 1 | \right],
\end{eqnarray}
on which Bob makes unitary transformation  $U = |0\rangle\langle 0| \pm |1\rangle\langle 1|$.
The fidelity is
\begin{eqnarray}
{\cal F}_4 &=& \langle Q  | U^\dag \rho_{B4} U | Q \rangle \nonumber\\
 &=& \frac{1}{\eta_2^2} \left[ | \gamma |^4 \cos^2 {u_a} \cos^2 {u_b} + | \delta |^4 \left( 1 + \sin^2 {u_a} \sin^2{u_b} \right)  \right.\nonumber\\
 && + \left. | \gamma  |^2 | \delta |^2 \left( \sin^2 {u_a} \cos^2{u_b} + \cos^2 {u_a} \sin^2 {u_b} + 2 \cos {u_a} \cos {u_b} \right) \right].
\end{eqnarray}

If the Bell measurement results in  $|\phi^-\rangle$,  with probability $p_2\equiv (|\delta|^2+ |\delta|^2 \sin^2 u_a +|\gamma|^2 \cos ^2 u_a)  /4$, the  state  of whole system is
\begin{eqnarray}
| {\phi^{QAB}_4}' \rangle &=& \frac{1}{\eta_2} | \phi^\pm_{QA} \rangle \left(- \gamma \cos {u_a} \cos {u_b} | 0 \rangle_{\rm{I}} | 00 \rangle_{\rm{II}}- \gamma \cos {u_a} \sin {u_b} | 1 \rangle_{\rm{I}} | 01 \rangle_{\rm{II}} \right.\nonumber\\
&&+ \left. \delta \sin {u_a} \cos {u_b} | 0 \rangle_{\rm{I}} | 10 \rangle_{\rm{II}} + \delta \sin {u_a} \sin {u_b} | 1 \rangle_{\rm{I}} | 11 \rangle_{\rm{II}} \pm  \delta | 1 \rangle_{\rm{I}} | 00 \rangle_{\rm{II}} \right). \end{eqnarray} The reduced density matrix observed by Bob is
\begin{eqnarray}
 \rho_{B4}' &=& \frac{1}{\eta_2^2} \left[ \left( | \gamma |^2 \cos^2 {u_a} \cos^2 {u_b} + | \delta |^2 \sin^2 {u_a} \cos^2 {u_b} \right) | 0 \rangle \langle 0 | \right) \nonumber\\
 && \mp \gamma \delta^* \cos {u_a} \cos {u_b} | 0 \rangle \langle 1 | \mp \gamma ^* \delta \cos {u_a} \cos {u_b} | 1 \rangle \langle 0 |\nonumber\\
 && + \left. \left( | \gamma |^2 \cos^2 {u_a} \sin^2 {u_b} + | \delta |^2 \sin^2 {u_a} \sin^2 {u_b}  + | \delta |^2 \right) | 1 \rangle \langle 1 | \right],
\end{eqnarray}
on which Bob makes unitary transformation $U = -|0\rangle\langle 0| \pm |1\rangle\langle 1|$.
The fidelity is
\begin{eqnarray}
{\cal F}_4' &=& \langle Q  | U^\dag \rho_{B4} U | Q \rangle = {\cal F}_4.
\end{eqnarray}

The average fidelity is
\begin{eqnarray}
{\cal F}_{\phi}&=& 2p_1{\cal F}_3+2p_2{\cal F}_4\nonumber\\ 
&=& \frac{1}{2}\left[ 1+ \sin^2 u_a \sin^2 u_b + \cos^2 u_a \cos^2 u_b + 2|\gamma|^2|\delta|^2(\sin^2 u_a \cos^2 u_b + \cos^2 u_a \sin^2 u_b  \right.\nonumber\\
&&\left. - \sin^2 u_a \sin^2 u_b - \cos^2 u_a \cos^2 u_b+2\cos u_a \cos u_b-1)\right],
\end{eqnarray}
which is symmetric between $\gamma$ and $\delta$.

\begin{figure}
\centering
\includegraphics[width=0.5\textwidth]{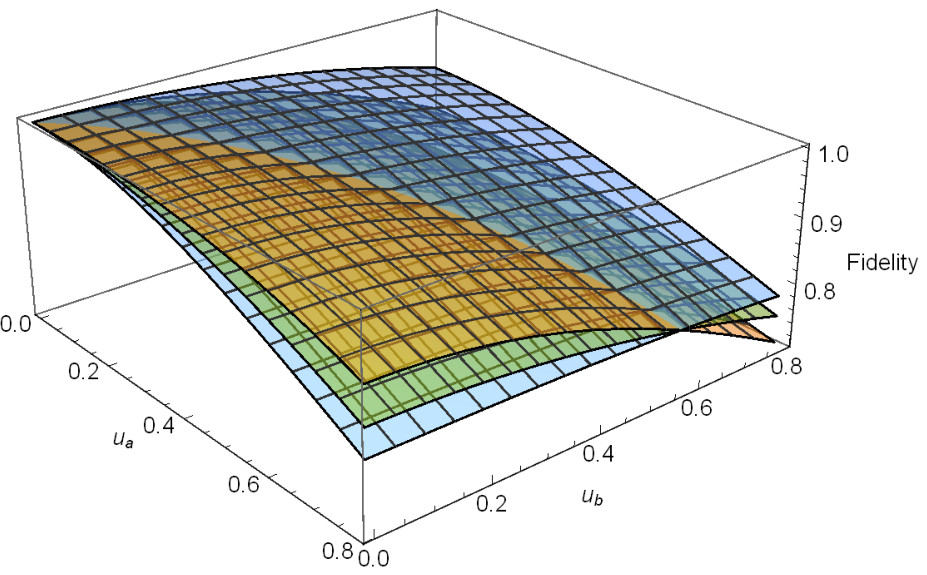}
\caption{\label{fid2}} For teleportation using $|\phi^\pm\rangle$,  the fidelities ${\cal F}_3$ (the yellow layer),  ${\cal F}_4$ (the blue layer) and ${\cal F}_\phi$ (the green layer) as functions of $u_a$ and $u_b$.  The  coefficients characterizing  the teleported state are   $|\gamma| = 0.6$ and $|\delta| = 0.8$.
\end{figure}

In general ${\cal F}_3 \neq {\cal F}_4$. Which is larger depends on the accelerations, unlike the relation between ${\cal F}_1$ and ${\cal F}_2$, which only depends on $|\gamma|^2$ and  $|\delta|^2$. For  $|\gamma|=0.6$ and $|\delta|=0.8$, we show ${\cal F}_3$, ${\cal F}_4$   and ${\cal F}_\phi$ in Fig.~\ref{fid2}.

\begin{figure}
\centering
\includegraphics[width=0.5\textwidth]{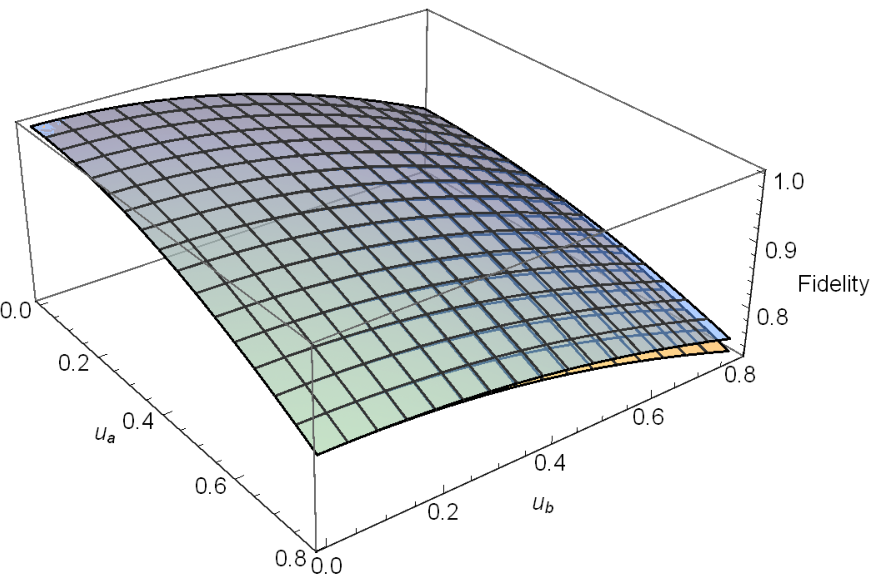}
\caption{\label{avefid}} Comparison between average fidelities ${\cal F}_\psi$ (the yellow layer) and ${\cal F}_\phi$  (the green layer) with  $|\gamma| = 0.6$ and $|\delta| = 0.8$.
\end{figure}

For  $|\gamma|=0.6$ and $|\delta|=0.8$, we show average fidelities ${\cal F}_\psi$ and ${\cal F}_\phi$  in Fig.~\ref{avefid}. It shows that for such a teleported state $|Q\rangle$, no matter what the accelerations are,    ${\cal F}_\phi > {\cal F}_\psi $, hence it is better to use $|\phi^\pm\rangle$.

When $|\gamma|^2=|\delta|^2=\frac{1}{2}$, ${\cal F}_1={\cal F}_2 ={\cal F}_3 = {\cal F}_4 = {\cal F}_\phi= \frac{1}{2} \left[ {1 + \cos {u_a} \cos {u_b}} \right]$.

\begin{figure}
\centering
\includegraphics[width=0.5\textwidth]{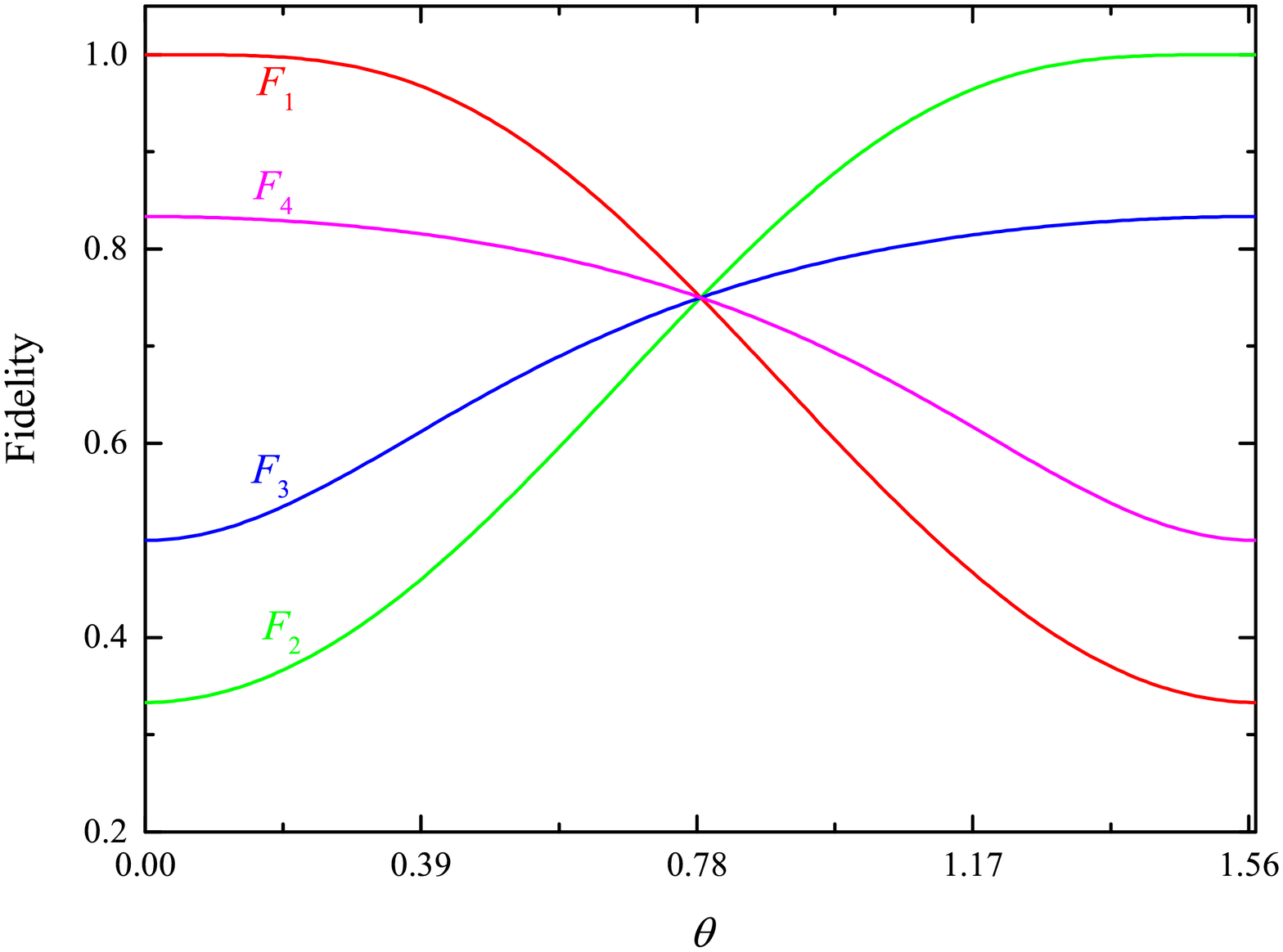}
\caption{\label{fidinf}} The dependence of the four fidelities on the state of the qubit when the accelerations are both observers are infinite, or  $u_a = u_b = \pi/4$. The parameter $\theta$ is defined as $\gamma = \sin \theta$ and $\delta = \cos \theta$.
\end{figure}

We now consider  the limit that the accelerations of both observers are infinite. In Fig. \ref{fidinf}, we show the dependence of each of the four basic fidelities on the coefficients characterizing the teleportated state.  We parameterize the coefficients as  $|\gamma| = \sin \theta$ and $|\delta| = \cos \theta$, with $0\leq \theta \leq \pi/2$. The four fidelities have a common crossing point at $\theta=\pi/4$, where  they are all equal, as we have known for general values of accelerations. Moreover, it can be seen that for any value of $\theta$, none of the four basic fidelities is zero in this limit of infinite accelerations.  In fact in the limit of infinite accelerations, the average fidelities  are
\begin{eqnarray}
{\cal F}_\psi &=&\frac{1}{2} + |\gamma|^2 |\delta|^2 \leq \frac{3}{4},\\
{\cal F}_\phi &= &\frac{3}{4}.
\end{eqnarray}
$1/2 \leq  {\cal F}_\psi \leq \frac{3}{4} $. Therefore, no matter what $|Q\rangle$ is, we always have ${\cal F}_\phi \geq {\cal F}_\phi$, hence it is better to use $|\phi^\pm\rangle$.

Now consider the special case that only Alice or Bob accelerates while the other observer  moves uniformly. If  Alice moves uniformly, then $u_a=0$, we have   ${\cal F}_1 = {\cal F}_4$, ${\cal F}_2 = {\cal F}_3$, $p_1=p_2=1/2$, thus ${\cal F}_\psi = {\cal F}_\phi$.    If Bob moves uniformly, then $u_b=0$, we also have ${\cal F}_1 = {\cal F}_4$, ${\cal F}_2 = {\cal F}_3$,  but $p_1\neq p_2$ in general, hence the difference between ${\cal F}_\psi$ and $ {\cal F}_\phi$  remains, unless $|\gamma|=|\delta|=1/\sqrt{2}$.

\section{Summary}\label{seccon}

We have studied how quantum entanglement between two Dirac modes is affected by the  accelerations of the observers of these two modes. Acceleration causes the observer to be  only able to  access a part of the spacetime. This is the origin of the entanglement degradation. We use negativity as the entanglement measure. For two kinds of entangled states, it is calculated that the nagativity decreases with the acceleration of each observer. However, there is a residual nonzero value even when both accelerations approach infinity.

Moreover, we have also studied how this entanglement degradation affects the quantum teleportation, by calculating the fidelities as functions of both accelerations.  In addition, the fidelity   depends on several factors, including which Bell state the observers share is $|\psi^\pm\rangle$ or $|\phi^\pm\rangle$, the result of the Bell measurement  by Alice, as well as  the state of the qubit to be teleported. We have obtained the fidelity for each case, as well as the average over all possible results of Bell measurement.   When the teleported state is an equal superposition of $|0\rangle$ and $|1\rangle$, all the fidelities are equal.

In the limit that both  accelerations are infinite, the average fidelity in using  $|\phi^\pm\rangle$ is $\frac{3}{4}$, which is higher than using $|\psi^\pm\rangle$, which is $\frac{1}{2} + |\gamma|^2 |\delta|^2\leq 3/4$. Therefore,  it is better to use $|\phi^\pm\rangle$.

For teleportation using boson modes, the fidelity approaches zero even when the acceleration of only one observer approaches infinity while the other observer moving uniformly \cite{alsing.teleportation}.  This demonstrates that the Dirac modes are advantageous over Boson modes in quantum teleportation in non-inertial frames.  

Accelerated observers   reflects some  situation in a gravitational field, because gravitation is equivalent to acceleration according to  Equivalence Principle. Furthermore, the spacetime near the horizon of a Schwarzschild black hole is the Rindler spacetime, and accelerated observers in the Minkowski spacetime correspond to stationary observers near a large black hole \cite{leonard2004introduction,rindler2012essential}. Our results suggest that Dirac modes are more useful than boson modes for quantum teleportation in a strong gravitational field, especially near a black hole.

This work was supported by National Science Foundation of China (Grant No. 11574054).

%\bibliography{Dirac}

\end{document}